# Users Authentication and Privacy control of RFID Card

Ikuesan R. Adeyemi  
raikuesan2@live.utm.my  

Norafida Bt. Ithnin  
afida@utm.my  

Department of Computer System and Communications,  
Faculty of Computer Science and Information Systems,  
Universiti Teknologi Malaysia.

**Abstract**— Security and Privacy concerns in Radio frequency identification (RFID) technology particularly RFID Card, is a wide research area which have attracted researchers for over a decade. Authenticating users at the Card end of the RFID technology constitutes one of the major sources of attacks on the system. In this research, we studied the various known attacks and mitigation available. We proposed a conceptual framework that that can be used to mitigate the unauthorized use of RFID Card. This concept will mitigate the single point of the RFID card failure: unauthorized use.

## 1 INTRODUCTION

Radio frequency identifier (RFID) is one among the series of wireless technology gaining faster and wider adoption in our today society due to its portability, mobility and flexibility of use. These unique characteristics constitute its wide adoption in various application such as conveyor systems, container and document identification, stock inventorying, transport tracking system, time sensitive application, self monitoring application such as expiry date alerts, and anti-counterfeiting of product; consequently, a replacement to the traditional way of object identification: Bar-Code. Hence, traditional way of identification is fast paving way for the RFID agile mode of identification, which in itself has various security and privacy concerns, which is due to the inherent vulnerabilities in the architecture of the RFID system.

## 2 RFID ARCHITECTURE

The RFID system comprises the tag, reader, backend database, and or control unit as illustrated in Figure 1. The tag comprises a radio frequency chip, encoding and decoding circuitry, antenna unit, and or a memory unit. Depending on power capacity, a tag can be classified into passive, semi-active or active tag.

Tags without internal power supply, are called passive tags, tags without internal power supply but only uses the internal supply for its internal memory circuitry are called semi-active, while tags that uses its internal power unit to power both its internal circuitry and the antenna unit for communication are called active tags. Additionally, tags can be categorized based on their frequency of communication. The communication frequency between the tag and the reader determines the energy and read range, and in some instance, the size of the tag. The reader communicates with the tag through tag interrogation,

## 3 OPERATING FREQUENCY OF RFID SYSTEM

RFID tag can be classified based on the frequency they operate. Generally, they are classified into three categories namely; the low frequency (LF), high frequency (HF), and the ultra high frequency (UHF). The LF class of tags primarily operates at 125 kHz, and within the range ≥30 kHz and ≤300 kHz. The HF RFID tag (which is the commonly used tag) operates primarily at 13. 56 MHz, and within the range of ≤3 MHz and ≥30MHz. This is similar to the Federal communications commission (FCC) stated boundary of 13. 56 MHz +/-17 KHz Industrial Scientific and Medical (ISM) use [2].

The UHF class of RFID tags range between 300MHz to 3GHz, albeit, the UHF tags in the Gen-2 protocols operates in ranges of ≥866MHz and ≤960MHz but it applicability varies in different countries [1]. These operating frequencies determine the data rate and the read range of the RFID system. Based on frequency range, the operating read range of the passive RFID tag operates at ≥30cm, ≥1m, and ≥7m for LF, HF and UHF tags respectively. However, for an active tag, the nominal range could span 100m since it does not require the reader to power its internal circuitry[3].

## 4 APPLICATION OF RFID

Due to the extensive read range of the RFID tag, its application area has grown widely and yet RFID technology has not achieved its optimum application areas.

### 4.1 Contactless Payment System

This is the process of paying or transacting using a cashless medium. The Exxon-Mobil speed pass employs RFID to speed customer through fuel purchase [4]. A passive RFID tag is mounted on the vehicle or attached to the key chain of the consumers, which is activated by a reader attached to the pump of the fueling station. The reader handshakes with the tag and reads the encrypted number. This number is then sent through the linking cable between the reader and the pump to a satellite receiver of the gas station. This is then sent to a datacenter where the authorization, verification and accounting are done. The E-Z pass toll system is similar to the speed pass. When a car enters the toll zone, the reader antenna in the zone activates the car-mounted tag. An encoded number is then communicated back to the reader, which is further transmitted through a secured channel to a back end database and control system.

### 4.2 Electronic Article Surveillance (EAS) System

One problem in the retail industry is dealing with product leaving the store without proper payment, be it intentional or not [5]. The idea behind the EAS system is the AES system is to limit shoplifting through the adoption of RFID system. Passive RFID tag is integrated into items, upon purchase, the tag is deactivated or notification is made for verification [6]. This is achieved with the aid of the reader setting up an interrogation zone. When any item carrying a tag passes the door/entrance/exit, an alarm or surveillance system is activated. This is then deactivated upon purchase [7,8].

### 4.3 Container Identification and Tracking

For container shipping application, a typical RFID device is operated in an ultrahigh frequency or microwave range (e. g. 900MHz or 2. 45GHz in the U. S. A. or 5. 8GHz in Europe). RFID recognition system in container is applied by mounting the tag on the container/item/pallet, through hanged read/write or write/read equipment installed in the forklift or handset or a handset read write equipment to recognize the dynamic information on the tag. The information read can be transmitted to a monitor system or control database. Every cargo unit has embedded RFID tag and all the information each tag is stored on a central control computer of the warehouse. The read/write equipment detects and reports information about every cargo and automobile working and which cargo is transported [9].

### 4.4 E-Passport and Document Identification

The data-page of machine-readable passport is embedded with 44 characters bearing the name of the holder, country and passport number. The integration of RFID into machine readable document (MRD) therefore balances the need for electronic data storage with automated document control [10]. The e-passport contains a contactless RFID chip and an aerial embedded on one of its pages [11] and its cover has a built-in metallic shield/Faraday cage to prevent unauthorized reading of the tag inside the passport. The chip contains an operating system (OS), application program and a set of data grouping conforming to the international civil aviation organization (ICAO) logical data structure[6].

### 4.5 Tag Implantation

"Mu-chip" (the smallest passive transponder about 0. 4mm) can be embedded into a paper sheets to track document which can only be read at a range of a few centimeters due to the size of the antenna. Another smaller size of RFID tag is the Veri-chip. It is about the size of a grain of rice and it is often implanted into human being, pets, as well as live stocks [12]Implantation chips can also work in identifying wandering Alzheimer's patients who got out without identity or cognizance of their location and destination. On October 14, 2004 [13], an article titled "identity chip planted under skin approved for use in health care" was on the New York Times and much other publication. Veri-chip was cloned in less than ten minutes by a Canadian hardware developer for the purpose of an article in Wired magazine [14]. According to [15], 900 hospitals have agreed to participate in the Veri-Med system, and about 600 people have received implant while the company has begun direct-to-customer campaign in targeted market such as South Florida[16].

### 4.6 Contactless Smart Card

This area of RFID application has gained wider adoption in physical access control system, and cashless payment processes. Example of such includes toll payment, e-passport, building access control, and so on. In 1994 and 1995, around 1million of contactless smart Cards were produced per year for public transport application, the volume rose to 4million per year in 1996 and 1997 [17]. In building access control, RFID tag can simply be called a key for access [3]. This can be seen in proximity and vicinity Cards

### 5. SECURITY AND PRIVACY CHALLENGES IN RFID SYSTEM

This section presents the various challenges associated with the RFID system with respect to the physical layer.

### 5.1 RFID Tag Cloning Attack

Cloning or counterfeiting of tag is simply forging the data illicitly gathered from an authentic tag into another tag usually a blank tag. Cloning of tag is one of the possible end products of skimming attack [7],[18] ; an attack which harnesses the vulnerability of cheaper cost of production of tag[19].

### 5.2 Physical Attack

A close look at the architecture from the attacker perspective will reveal the delicate location of the antenna and even the RFID tag itself [6]. Depending on the intention of the attacker, physical attack could range from trying to explore manufacturer product, to complete destruction of tag. For instance, placing a tag inside in microwave oven is a direct way of frying a tag [7] while a simple EMP practice could damage the internal circuitry and even the tag as well.

### 5.3 Skimming Attack

Skimming attack is done by surreptitiously reading [6] the data of the tag without the authorization of the tag holder. This attack exploits the promiscuity of an RFID tag. An RFID tag does not have the intelligence to decide when to function or not, and so, it continually transmit beacons to any reader available to it. Additionally, it is a major tool in exploiting the vulnerabilities in RFID technology [11].

### 5.4 Spoofing Attack
Spoofing attack is an impersonation attack in which an attacker masquerades as a reader in order to interrogate tags [6]. The obtained response is then used as a means for later communication. When the legitimate reader queries the tag, the attacker will send the obtained response to the reader [18],[20].

### 5.5 Relay Attack

This attack can also be referred to as a Man-in-the-middle (MITM) attack. It is a technical way of fooling RFID tag and reader, as though they are communicating with each other. The attacker devices two units; a malicious reader which is called a Mole and a malicious tag called Proxy and both are connected via a communication link. The mole is set to interface with the authentic tag while the proxy interfaces with the reader. The Mole initiates communication with the tag, and then sends the response to the proxy. The proxy then forwards the response to the authentic reader. The fooled authentic reader sends back its response to proxy, which forwards it to the authentic tag through the Mole[21],[22]. [23] demonstrated a practical relay attack on e-voting system. similarly, [19] performed a practical relay attack on the communication process between a tag and a reader.

### 5.6 Denial of Service (DOS) Attack

His attack involves jamming the operating frequency of the communication channel of the RFID system: compromising the availability of resources. It could be denial of reading or denial of authentication [24], tag destruction [6]. This type of attack is aimed at preventing communication between the tag and the reader.

### 5.7 Clandestine Tracking
This is an illicit acquisition of information usually for the purpose of subversion or deception. This is major privacy concern [25]. Clandestine tracking exploits the ubiquitous nature of the RFID tag, thus making it a very common and easy-to-achieve attack. Additionally, the EPC network does not provide explicit protection against clandestine tracking [26].

Various researchers have studied these vulnerabilities, and proposed numerous protocols, methods, technique as well as standards against these attacks.

## 6. RELATED RESEARCH

The adoption of RFID technology into areas like physical access control have generated questions such as; 'how can I know when my Card is being read', how can the Card detect the authentic owner of the Card. Series of such question have trailed the stage of RFID technology

### 6.1 RFID Guardian

RFID guardian addressed issues such as denial of service, and privacy concerns while still enhancing the prevention of RFID ubiquity vulnerability exploitation through the integration of query audit, key management, access control and authentication [5]. It is a portable battery powered device that mediates interaction through selective frequency jamming and spoofing between RFID reader and a transponder [20] capable of two-way communication [5]. RFID Guardian establishes a privacy zone around the user in which only authenticated readers are allowed access, by acting like a reader, querying tag and decoding tag response [6].

However, one main disadvantage accrued to RFID guardian is range. Since it is expected to guard all tag in the user vicinity, its range should be between 1-2m [6]. This is a violation of ISO 14443-reader range specification of 10cm. In addition, the guardian itself represents a single point of failure; if the guardian fails or compromises, the user is unprotected and it can be easily lost or even forgotten [6]. Integrating RFID guardian into a PDA (for instance) could reduce this weakness but not prevent it. Being a battery powered device, an attack could target draining of its power by flooding the communication zone with series of irrelevant communications. Furthermore, it does not guarantee an unauthorized reader from knowing the existence of the tag.

### 6.2 RFID Blocker Tag

Blocker tag [27] simulates the full sets of $2^k$ possible RFID-tag serial number. It has the ability to block the RFID tree-walking singulation algorithm [7],[28] protocol used by the reader to select a particular tag. The blocker tag simulates the full spectrum of the possible serial number of the tag thereby obscuring the serial number of other tags. Blocker tag may be used to establish a save zone around a tag [6], preventing tag from being read. In order to make blocker tag more flexible, it is possible to implement a form of selective blocking [6]. Hence, a blocker tag helps to protect the privacy of the user from a malicious hidden reader.

The principal point of weakness of the blocker tag is the lack of flexibility [6]. Reliance on a blocker tag in a sparsely populated tag and reader environment does not prevent tracking of the tag and even active jamming. Blocker tag therefore provides temporary solution to privacy concerns and lesser or no security solution to RFID tag in the RFID system.

### 6.3 Labeling
Labeling a tag or content containing RFID tag is one way of informing a user of the presence of the RFID tag [18]. It is a major awareness ground upon which privacy of RFID system is elucidated, and accepted as part of the Bill-of-Right [6].

However, it also announces itself to malicious user. Hence, it complicates issues for security unravelment. Although, it works for privacy concerns but aids the exploitation of the vulnerability of the system by an attacker.

### 6.4 Kill Command

Some RFID tag has built-in kill command [6]. Auto-ID center and EPC global created a kill-command specification for permanent tag inoperability [29] some of which requires 32-bit password; a Class-1Gen-2 EPC standard tag. Thus, a tag can be destroyed or killed by sending a special kill-command and including the right coded password. The inclusion of the password is to prevent unauthorized killing [6] as well as enforcing user confidentiality [30]. However, it also exposes the tag to unauthorized killing by a malicious adversary; hence, a built-in kill command can be used by a malicious to cause a denial of service of attack by inputting the correct kill-command, at an inappropriate time. Similarly, an authentic user can mistakenly input the correct built-in kill command at a wrong timing.

**6.5 RFID Zapper**

This is another method of enhancing the privacy in RFID technology. An RFID zapper practically creates an electromagnetic impulse (EMP) within the boundary of its influence [31],[18]. [31] demonstrated a practical Zapper with a set-up EMP from a low cost disposable camera. Zapper can destroy the tag without necessarily altering the product carrying the tag [6] as against a microwave oven. Additionally, zapper is portable, hence can enable destruction of tag at the point of purchase of goods.

**6.6 Authentication Protocols**

[32] proposed an authentication protocol on new HMAC-based protocol as advancement on the ones proposed by [33], [34] and [35]. The new Hash-based Message Authentication Code (HMAC-BASED) protocol in which H was assumed to be a one-way hash function [32] could also be used to prevent skimming attack. BurrowAbadiNeedham89 (BAN) logic [36] was adopted for this new HMAC-based protocol. RFID distance bounding protocol proposed by [21] can be effective in the defense against relay attack. According [21], relay attack introduces some delay in the transmission which conventional cryptographic measure employed in RFID system at the application layer due to much synchronization, and other mechanism for error-handling, anti-collision, e. t. c. cannot detect. [37] proposed a distance-bounding or secure-position protocol integrated into the physical layer of the communication protocol which can detect relay attack through high resolution timing information on bit arrival. Researches available in authentication of RFID systems are targeted at the logical mitigation leaving the physical authentication vulnerable to attacks.

**6.7 Physical-Layer Identification Technique**

The physical layer communication of the RFID system is the first layer of communication of the RFID system. It is also the gateway for RFID technology as well as the first point of call for security defenses and privacy regulation. [42] designed, implemented, and evaluated a technique called PARADIS: to identify the source network interface Card of an 802. 11 frame by analyzing the physical layer of the passive radio-frequency analysis. Radiometric identification (radio frequency fingerprinting) was the technique adopted by them for wireless device identification. Experimental result for physical identification by PARADIS yielded 99% accuracy [42]. [43] adopted same principle in their experiment on RFID passive UHF tag identification but utilizes time domain features and spectral principal component analysis (PCA) for extraction and matching of the fingerprinting respectively. Using similar principle, [44] experimented on physical-layer identification of RFID passive tags but on HF tags. Results from their experiments showed that RFID cloning challenge can be solved.

**6.8 Anti-Counterfeiting Technology**

The unique identifier of RFID systems is susceptible to forgery/cloning attack. This attack can be mitigated via challenge response authentication protocol [29]. Engineering Researchers at the University of Arkansas developed an anti-counterfeiting method against RFID tag [39]. In their research, they discovered that each tag has a minimum power response at multiple frequencies, constituting its unique physical characteristics. [41] proposed a chaos theory for detecting cloned tag. [40] demonstrated a resistive technique against cloning of RFID tags through challenge response authentication.

**6.9 Fingerprint Biometric Authentication on Smart Card**

[45] modeled a framework for such an authentication procedure, on a smart Card. In their framework, a fingerprint sensor was used in capturing the fingerprint. They reduced the captured image into minutiae points, and was stored as a template in the Card through a Card Reader. The authenticating process involved a match-on-Card (MOC), through a system-on-Card (SOC). Similarly, [46] proposed a two-factor authentication system based on combined fingerprint recognition and smart RF Card verification. From their experiment, they concluded that the optimization does not reduce the data precision, and that fingerprint recognition algorithm has good performance on hardware system.

**6.10 Controllable Tag**

[47] proposed a tag, which can be controlled by the user, sampling various experimental designs. In one sample, the tag was equipped with indicator (both acoustic and visual effect) to indicate tag in-use or otherwise. In order to control the activity of the tag, they integrated a switch system, which can be manually activated. Additionally, they designed tag that operates based on orientation, daylight and darkness, and proximity dependant. Similarly, [48] investigated the integration of a smart Card, and electrical switch to enhance high level of security and privacy protection measure. In their design, parallel plate conductors was used to decide the state of the switch, controlled with human finger. Clip tag [49] protects against privacy concerns in addition to post point of sale negotiation [6]. It does not provide any security guide or measure. Faraday cage/shield [50] can prevent tags from being read [6].

In view of the above, various proposed mitigation measure have not taken into consideration the vulnerabilities posed by user-end authentication at the physical-layer of the RFID technology. Consequentially, unauthorized use of tag has led to practical security and privacy breach in the
RFID system.

TABLE 1
RFID PHYSICAL-LAYER ATTACK-MITIGATION ANALYSIS

| Mitigation \ Attacks | Unauthorized killing | Unauthorized reading | Unauthorized use | Cloning | Skimming | Spoofing | Clandestine tracking | Relay | Privacy |
|---|---|---|---|---|---|---|---|---|---|
| Clipped Tag | ö | ö | ö | × | ö | × | √ | ö | √ |
| Faraday Cage | √ | √ | × | × | √ | × | √ | ö | ö |
| Controllable Tag | √ | √ | ö | × | √ | × | √ | ö | √ |
| Biometric authentication | ö | ö | √ | ö | ö | ö | ö | ö | ö |
| Anti-counterfeiting | × | × | × | √ | × | √ | × | √ | × |
| Physical-layer identification technique | × | × | × | √ | × | √ | × | √ | × |
| Authentication protocol | ö | ö | × | ö | √ | ö | ö | ö | ö |
| Labeling | Creates user awareness ||||||||
| RFID Guardian | ö | √ | × | × | √ | × | √ | ö | ö |
| RFID Blocker Tag | √ | √ | × | × | √ | × | ö | ö | √ |

√ indicated that the mitigation measure is applied to such threat
× indicates that the measure does not apply to that threat
ö indicates the measure can be combined with other measures

Unauthorized tag use/read particularly at the tag end can be mitigated through the integration of methods such as clip tag, biometric authentication protocol, and Faraday shield as indicated in Table 1.

**7. PROPOSED FRAMEWORK**

From the analysis of Table 1, we proposed a conceptual framework which can be integrated into the RFID Card to mitigate the unauthorized use of Card.

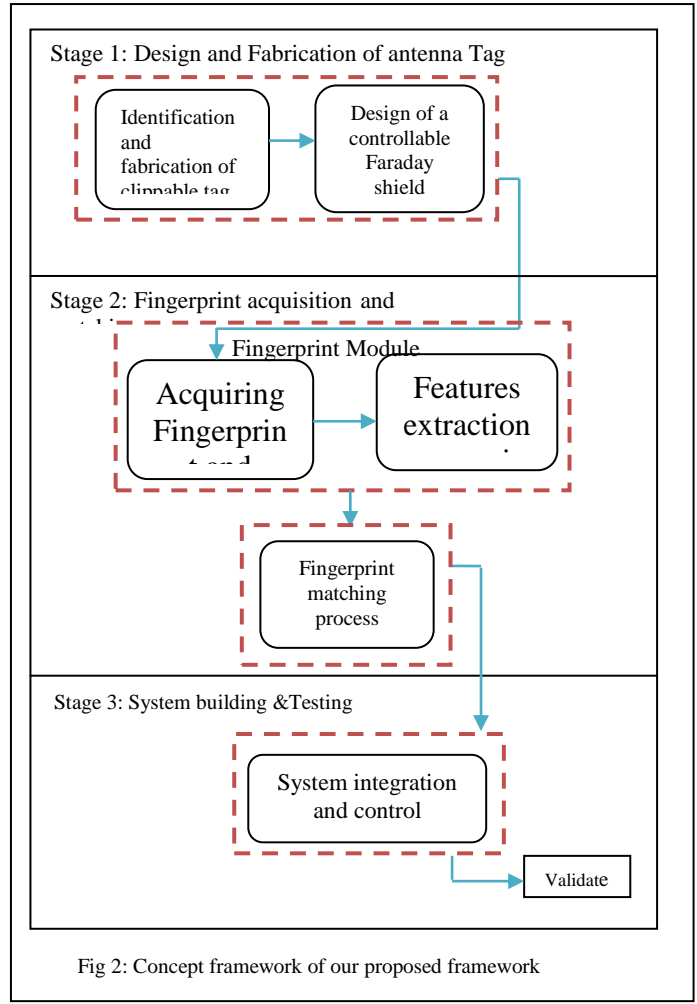

Fig 2: Concept framework of our proposed framework

Our research aim to integrate biometric authentication process and a Faraday shield into a clippable RFID Card. The processes are divided into three stages as shown in figure 2.

*Stage 1*: Design and fabrication of tag
This stage comprises the design, calibration, simulation and fabrication of the tag antenna, and a controllable joint connected to a controller.

*Stage2*: Fingerprint Matching and storage
This stage involves the process of acquiring, authenticating, securing and storage of the biometric authentication process, fingerprint in this case.

*Stage3*: System building and testing
This stage entails the integration of the various phases in a secured control unit and testing of the prototype.

**CONCLUSION**

In this research, we identified the various challenges facing RFID Card, with careful reference to the physical layer. We also highlighted the diverse application areas of the RFID technology as well as some possible area of application. Furthermore, we proposed a conceptual framework to mitigate the challenges facing RFID card; unauthorized use of Card in particular. This concept will be fully implemented in our future work, and will be evaluated in line with the various known attacks and strategies.